\documentclass[showpacs,preprint2,amsmath,amssymb]{revtex4}
\usepackage{graphicx}
\usepackage{dcolumn}
\usepackage{subfigure}
\usepackage{epsfig}
\usepackage{bm}
\usepackage{caption}
\newcommand{\barr}{\begin{eqnarray}}
\newcommand{\earr}{\end{eqnarray}}

\newcommand\Tstrut{\rule{0pt}{2.6ex}}  
\newcommand\Bstrut{\rule[-1.2ex]{0pt}{0pt}}
\newcommand{\dis}{\displaystyle}

\usepackage{hyperref}

\begin{document}


\title{Supernova neutrino scattering
off Gadolinium even isotopes in water Cherenkov detectors.}

\author{Paraskevi C. Divari}
\affiliation{Department of Physical Sciences and Applications,
Hellenic Military Academy, Vari 16673, Attica, Greece}

\begin{abstract}
Neutrinos in water can be detected thanks to several reactions.
The most important one is the inverse beta decay $\bar\nu_e +p
\rightarrow n+e^+$. The detection of 2.2 MeV $\gamma$ from neutron
capture on free protons is very difficult. The feasibility of
Gadolinium (Gd) doping in water Cherenkov detectors essentially
reduces background signals and enhances the sensitivity to
neutrino detection. In this work  the supernova neutrino
charged-current interactions with   the most abundant Gd even
isotopes (A=156,158 and 160) are studied.   We use measured
spectra and the quasiparticle random phase approximation   to
calculate the charged current response of Gd isotopes to supernova
neutrinos. Flux-averaged cross sections are obtained considering
quasi-thermal neutrino spectra.
\end{abstract}
 \maketitle

\section{ Introduction}

A number of detectors like water Cherenkov detectors (WCDs)
\cite{IMB,SK,SNO}, have been used in various neutrino detection
experiments.
They have the ability to detect either the charged-current
${\nu}_{e}$ ($\bar{\nu}_{e}$) interaction, which produces
electrons (positrons), or the neutral current interaction (for all
flavors), which usually results in the production of neutrons and
photons, or both. The sensitivity of the detectors can be enhanced
through either building a larger water tank that increases the
probability of neutrino interaction in WCD,
 or including additives, such as
gadolinium (Gd), in water that essentially reduces background
signals \cite{Laha}. Neutrinos in water can be detected thanks to
several reactions.
The most important  are the following three:
\begin{enumerate}
    \item inverse beta decay (IBD):
     $\mathrm{p} + \overline{\nu}_{\mathrm{e}}\rightarrow \mathrm{n} + \mathrm{e}^+$
    \item elastic scattering on electrons :
     $\nu + \mathrm{e}^-\rightarrow \nu + \mathrm{e}^-$
    \item neutral current scattering on oxygen :
     $\nu + \text{\textsuperscript{16}O}
    \rightarrow \nu+\text{\textsuperscript{16}O}^*$
\end{enumerate}
 with $\nu={\nu_e,\bar\nu_e,\nu_{\mu,\tau},\bar\nu_{\mu,\tau}}$

 WCDs are   primarily sensitive to $\bar{\nu}_{e}$'s through
IBD. The positron produced by IBD emits Cherenkov light, which is
detected by a photomultiplier tube (PMT) array placed around the
detection volume. Due to Cherenkov threshold the detection of the
2.2 MeV $\gamma$  from neutron capture on hydrogen nucleus
($n+p\rightarrow d + \gamma\hspace{2pt} (2.2\hspace{1pt}MeV$ )) is
very difficult. It is possible to dissolve Gd compounds in the
water to enhance neutron tagging and allow the IBD and electron
elastic scattering signals to be separated \cite{Laha,Vagins}. The
large neutron capture cross section of Gd allows neutrons formed
in IBD events to be quickly ($\sim 20 £\mu s$) captured, emitting
three to four gamma rays with a total energy of 8 MeV
($n+\text{Gd}
  \rightarrow
\text{Gd\textsuperscript{*}} + \gamma\hspace{2pt}
(8\hspace{1pt}MeV$)) in close time and space
 coincidence with the positron. In Super-Kamiokande(SK),
 which is a 32~ktons (fiducial) WCD,
 it has be found that  the inclusion of $\text{GdCl}{_3}$ salt  (0.2\% in weight)  to
SK, $\sim 90\%$ of the IBD events could be tagged
\cite{Laha,Vagins}. The remaining IBD events  as well as the
$\bar{\nu}_{e}$ absorption events on \text{\textsuperscript{16}O}
can then be statistically subtracted from the remaining signal.


Future extremely large WCDs like Hyper-Kamiokande (560~ktons
fiducial) would have a dramatic impact on detecting supernova or
solar neutrinos using the Gd-doping technique. Therefore, it would
be interesting to draw our attention to the possibility of
calculating the cross sections for low-energy neutrinos on Gd
isotopes. In the present work we pay special attention on
calculations of charged current (CC) neutrino/antineutrino-Gd
cross sections at neutrino energies below 100 MeV, considering the
most abundant even isotopes of Gadolinium that is, isotopes with
mass number  A=156,158 and 160 (20.47\%, 24.84\% and 21.86\%
abundant, respectively).
The corresponding nuclear matrix elements have
been calculated in the framework of quasi-particle random phase
approximation(QRPA) \cite{Divarijpg2,Divarixe132,DivariINTECH}.


\begin{figure}[htb]
\begin{center}
\hspace{1cm}\includegraphics[scale=0.6]{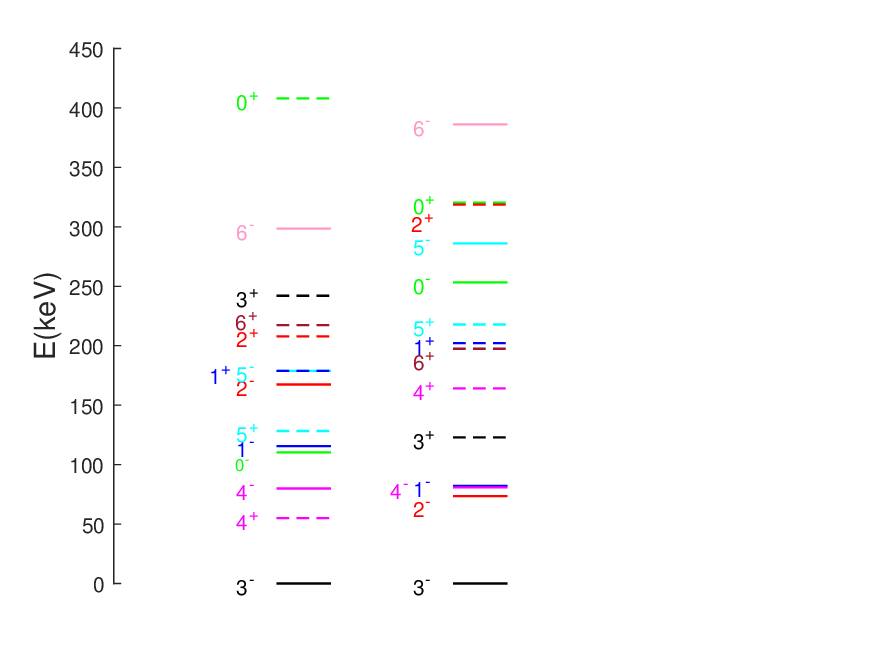}
\caption{(Color on line)  Experimental (left)\cite{tb158} and
theoretical (right) spectra of $^{158}$\hspace{1.0pt}Tb}
\label{fasma}
\end{center}
\end{figure}

\section{Brief description of the formalism}

The   standard model effective Hamiltonian in  the  charged
current reactions
\begin{eqnarray} \label{ch}
(A,Z)+ {\nu}_{e}\rightarrow (A,Z+1)+e^-\nonumber\\
(A,Z)+\bar{\nu}_{e}\rightarrow (A,Z-1)+e^+\nonumber
\end{eqnarray}
can be    written
\begin{equation} \label{hamil}
{\cal H} \, =\, \frac{G_F \hspace{3pt}cos\theta_c}{\sqrt{2}}
j_{\mu} ({\bf x}) J^{\mu} ({\bf x}),
 \end{equation}
A(Z) represents the mass(proton) number of a nucleus,
respectively. Here $G_F =1.1664\times 10^{-5}$$ GeV^{-2}$ denotes
the Fermi weak coupling constant and $\theta_c\simeq 13^o$ is the
Cabibbo angle. According to V-A theory, the leptonic current takes
the form \cite{Walecka,Donnelly,Donnelly1,Donnelly2}
\begin{equation}\label{lcurrent}
j_\mu = \bar{\psi}_{\nu_{\ell}}(x)\gamma_\mu (1 - \gamma_5)
 \psi_{\nu_{\ell}}(x)\, ,
\end{equation}
where $\psi_{\nu_{\ell}}$ are the neutrino/antineutrino spinors.
 The hadronic current of vector, axial-vector and
pseudo-scalar components  is written as
 \begin{eqnarray} \label{hcurrent}
J_{\mu}=\bar{\Psi}_N\big[ F_1 (q^2)\gamma_{\mu}+F_2 (q^2)\frac{i
\sigma_{\mu \nu}q^{\nu}}{2M_N}+F_A (q^2)\gamma_{\mu}\gamma_5 \nonumber\\
+F_P(q^2)\frac{1}{2M_N}q_{\mu}\gamma_5 \big]\Psi_{N}
\end{eqnarray}
($M_N$ stands for the nucleon mass, $\Psi_{N}$ denotes the nucleon
spinors and  $q^2$, the square of the four-momentum transfer). By
the conservation of the vector current (CVC), the vector form
factors  $F_{1,2}(q^2)$ can be written in terms of the proton and
neutron electromagnetic form factors \cite{athar06}.
 The axial-vector form factor $F_A(q^2)$ is
assumed to be of dipole form  \cite{Singh} while   the
pseudoscalar form factor $F_P(q^2)$ is obtained from the
Goldberger-Treiman relation~ \cite{Walecka}.

In the convention we used in the present work   the square of the
momentum transfer, is written as
\begin{equation} \label{qeq} q^2=q^{\mu}q_{\mu} =\omega^2- {\bf q^2} =
(\varepsilon_i - \varepsilon_f)^2-({\bf p}_i-{\bf p}_f)^2 \, ,
\end{equation}
where $\omega=\varepsilon_i-\varepsilon_f$ is the excitation
energy of the final nucleus. $\varepsilon_i$(${\bf p}_i$) denotes
the energy(3-momenta) of the incoming neutrino/antineutrino and
$\varepsilon_f$(${\bf p}_f$)  those of the outgoing
electron/positron,
 respectively.
The charged-current  neutrino/antineutrino-nucleus  cross section
is written as \cite{Donnelly}
\begin{eqnarray}
\label{eq:Sec2_1}
\sigma(\varepsilon_i)=& \dis{ \frac{2 G_F^2
cos^2\theta_c}{2J_i+1}} \sum_f
 {|{\bf p}}_f|\varepsilon_f\int_{-1}^{1}d(\cos{\theta})
 F(\varepsilon_f, Z_f)\nonumber\\
 &\times\big(\sum \limits_{J=0}^\infty \sigma_{CL}^J(\theta)
+
  \sum \limits_{J=1}^\infty \sigma_{T}^J(\theta) \big)
\end{eqnarray}
$\theta$  denotes the lepton scattering angle. The summations in
Eq. (\ref{eq:Sec2_1}) contain the contributions $\sigma_{CL}^J$,
for the Coulomb $\widehat{\mathcal{ M}}_J$ and longitudinal
$\widehat{\mathcal{ L}}_J $, and $\sigma_{T}^J$, for the
transverse electric $ \widehat{\mathcal{T}}_J^{el}$ and magnetic $
\widehat{\mathcal{T}}_J^{mag}$  multipole operators defined as in
Ref. \cite{Divarijpg2}. These operators include both polar-vector
and axial-vector weak interaction components.

\begin{figure}[htb]
\begin{center}
\hspace{1cm}\includegraphics[width=0.45\textwidth]{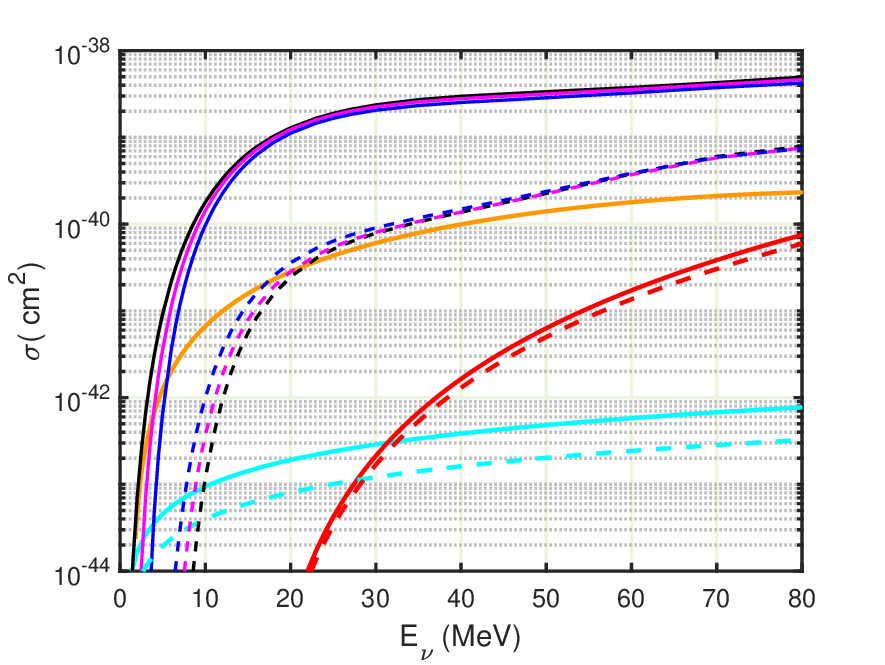}a)
\includegraphics[width=0.45\textwidth]{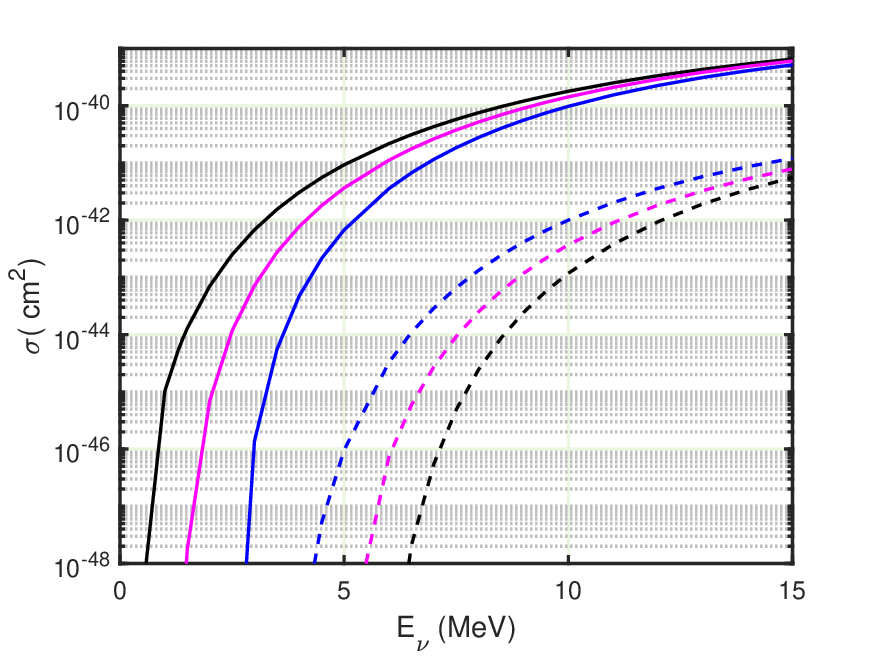}b)
\includegraphics[width=0.48\textwidth]{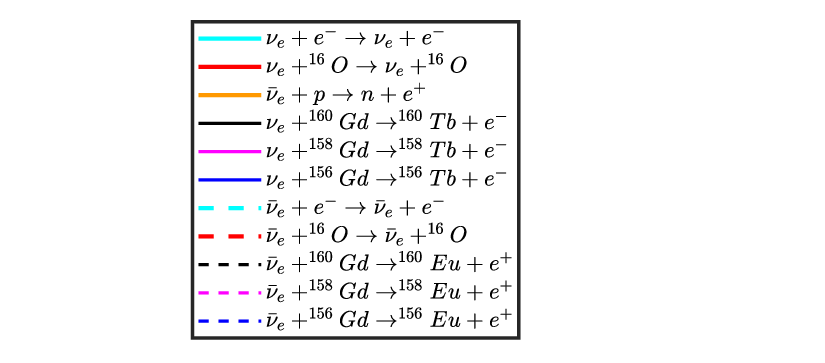}
\caption{ (Color on line) (a) Total cross sections for  $\nu_e$
and $\bar\nu_e$ Gd interactions. Also presented are  total cross
sections for inverse beta decay, elastic scattering on electrons
and neutral current scattering on oxygen. (b) Total cross sections
for $\nu_e$ and $\bar\nu_e$ Gd interactions in the energy region
from 0 to 15MeV. } \label{cs1}
\end{center}
\end{figure}

\section{Neutrino spectra}

 Energy distributions of supernova neutrinos are shaped by the
circumstances in which the neutrinos are emitted. Neutrinos
leaving the star are responsible for the cooling of the
proto-neutron star forming in the star's core. Hence, their
spectrum resembles a thermal one, with temperatures reflecting the
conditions at the site where they decoupled. However, the fact
that different kinds of neutrinos are involved in different
interactions, and that the reactivity of the neutrino/antineutrino
depends on its energy, flavor, and helicity, modulates this
picture. For all neutrino/antineutrino flavors, the energies are
in the range of a few to a few tens of MeV, although calculations
of neutrino transport that use different opacities achieve
somewhat different spectra.

There are different ways to characterize the spectra of the
neutrino time integrated fluxes emerging from a Supernova (SN).
 Recent results showed the
supernova-neutrino energy distribution to be accurately
parameterized with a power-law distribution
\cite{Lujan-Peschard:2014lta,Tamborra:2012ac}:
\begin{eqnarray}\label{pldis}
\eta_{PL}(E_\nu)=
    \frac{E_{\nu}^{\alpha_i}\:
    e^{-E_{\nu}/T_i}}{T_i^{\alpha_i +1 }\:
    \Gamma\left(\alpha_i+1\right)}
\end{eqnarray}
adopting the Keil parametrization \cite{Keil:2002in} for the
neutrino fluence
\begin{eqnarray}\label{eq:dFdE}
\mathcal{F}_i^0\left(E_{\nu}\right) =
    \frac{\mathrm{d} F_i^0}{\mathrm{d} E_{\nu}} =
    \Bigg(\frac{\mathcal{E}_i}{\langle E_i\rangle 4\pi D^2}\Bigg)\eta_{PL}(E_\nu)
\end{eqnarray}
with $i = \nu_{\mathrm{e}},\,
    \overline{\nu}_{\mathrm{e}},\,\nu_x$ ,
    $\nu_x=\nu_{\mu,\tau},\bar\nu_{\mu,\tau}$,
where   $E_{\nu}$ is the neutrino energy, $\Gamma(x)$ the Euler
gamma function, $T_i$ the temperature
\begin{equation}
    T_i = \frac{\langle E_i\rangle}{(\alpha_i + 1)},
    \label{eq:TEmedia}
\end{equation}
$\langle E_i \rangle$  being the mean energy and  $\alpha_i$ a
parameter called the pinching parameter that relates to the width
of the spectrum. Typically $\alpha_i$ takes the values $2.5-5$ for
time dependent flux \cite{Keil:2002in} depending on the flavor and
the phase of neutrino emission.  Eq. (\ref{eq:dFdE}) is observed
to be closer to thermal distribution than the time-dependent flux.
A reasonable conservative interval for $\alpha$
is\cite{Vissani:2014doa} $1.5 \le \alpha\le 3.5$. $\mathcal{E}_i$
denotes the total energy in that $i$ flavor and D is the distance
to the supernova. A SN at a distance $D=10$kpc emits total energy
$\approx 3\times10^{53}$ erg over a burst $\Delta t \approx 10$s
in neutrinos of all six flavors
\cite{Costantini:2005un,Mirizzi:2006xx,Adams:2013ana}. The
$\nu_x=\{\nu_{\mu,\tau},\bar\nu_{\mu,\tau}\}$ have similar
interactions and thus similar average energies and fluences.
Therefore, the total energy is divided as
$\mathcal{E}={\mathcal{E}}_{\nu_e}+{\mathcal{E}}_{\bar\nu_e}+4{\mathcal{E}}_{\nu_x}$.
In typical SN simulations the equipartition hypothesis  among the
primary flavors is taken
${\mathcal{E}}_{\nu_e}\approx{\mathcal{E}}_{\bar\nu_e}\approx
{\mathcal{E}}_{\nu_x}=5\times10^{52}\text{erg}$.

According to the simulations in \cite{Tamborra:2012ac}  and the
findings from the SN1987A\cite{Loredo02,Pagliaroli09}, the average
energy for the electron antineutrino can be set to $\langle
E_{\bar\nu_e}\rangle=12 \hspace{2pt}\text{MeV}$. The mean energy
of the non-electronic species $\nu_x$ can be taken 30\% higher
than the one of the $\bar\nu_e$ that is $\langle
E_{\nu_x}\rangle=15.6\hspace{2pt}\text{MeV}$ compatible with what
is found in \cite{Keil:2002in}. The electron neutrino mean energy
can be taken from the condition that the proton (or electron)
fraction of the iron core in the neutron star forming is 0.4 which
gives $\langle E_{\nu_e}\rangle=9.5\hspace{2pt}\text{MeV}$
\cite{Lujan-Peschard:2014lta}.

\section{Results}

The nuclear  matrix elements   entering in Eq. (\ref{eq:Sec2_1})
  have been  calculated in the framework of pnQRPA. The  target isotopes  $^{156,158,160}$Gd   were
assumed to be at the BCS ground state (initial state). The final
excited states $|J^{\pi}_f>$ of~ $^{156,158,160 }$Tb
($^{156,158,160}$Eu)   isotopes have been calculated by solving
the pnQRPA equations \cite{Divarijpg2}. The active model space for
protons  consists of the complete oscillator shells $4\hbar\omega$
and $5\hbar\omega$ while for neutrons the   oscillator shells
$5\hbar\omega$ and $6\hbar\omega$. The corresponding single
particle energies (s.p.e) were produced by the well known Coulomb
corrected Woods-Saxon potential adopting the parameters of Bohr
and Mottelson \cite{Bohr-Mot}. The quality of the obtained results
could be improved   adjusting some of the proton and neutron
single particle energies. These adjustments are presented in Table
\ref{adj}.
\begin{table}[htb]
 \caption{Adjusted (Adj) single-particle energies
  together with the Woods-Saxon (WS) energies   (in MeV) for
the neutron ($n$) and proton ($p$) orbitals.}
\begin{center}
\begin{tabular}{lclllllll}
\hline\Tstrut\Bstrut
orbital  & $^{156}$Gd    & &  $^{158}$Gd  & &  $^{160}$Gd  \\
\hline
                &  WS   & Adj    & WS &  Adj &WS &Adj \Tstrut\Bstrut \\
\hline
 $n \hspace{2pt}1\textrm{f}_{7/2}$ & -6.43 & -6.10  &-6.40  & -6.07 &-6.37 &-6.03 \\
$n\hspace{2pt} 0\textrm{h}_{9/2}$ & -5.68 & -5.00   &-5.71  & -5.03&-5.73 &-5.05 \\
$n\hspace{2pt} 0\textrm{h}_{11/2}$ & -11.05 & -6.00 &-10.99  &-5.94 &-10.93 & -5.87 \\
$p \hspace{2pt}0\textrm{h}_{11/2}$ & -5.43 & -5.00  &-6.01  &-5.58 &-6.58 &-6.14\\
 \hline
\end{tabular}
\end{center}
\label{adj}
\end{table}

The two-body  matrix elements were obtained from the Bonn
one-boson-exchange potential  applying the G-matrix techniques
\cite{Holinde}. Pairing interaction between the nucleons can be
adjusted by solving the BCS equations. Specifically, the monopole
matrix elements of the two-body interaction are scaled by the
pairing-strength parameters $g_{pair}^p$ (for protons) and
$g_{pair}^n$ (fot neutrons) in such a way that the
 resulting lowest
quasiparticle energy to reproduce the phenomenological pairing gap
$\Delta_{p,n}^{exp}$ \cite{Audi}.
In Table \ref{pairing}
 the values of the pairing-strength parameters, as well as
the theoretical energy gaps ($\Delta_{p,n}^{th}$)   determined at
the BCS level are tabulated. Also listed is the oscillator length
parameter $b$ for each isotope as well as their corresponding
natural abundances.
\begin{table}[tbp]
\par
\vskip 0.5cm \caption{Pairing-strength parameters for protons (
$g^p_ {pair}$) and neutrons ($g^n_ {pair}$) determined by solving
iteratively  the BCS equations. They are fixed in such a way that
the corresponding experimental energy gaps (in MeV) for protons
($\Delta_{p}^{exp}$)  and neutrons ($\Delta_{n}^{exp}$ )  to be
reproduced. The values of the harmonic oscillator size parameter
$b$ as well as the corresponding natural abundances for each
isotope   are also shown.}
\renewcommand{\tabcolsep}{0.5pc} 
\renewcommand{\arraystretch}{1.5} 
\begin{center}
\begin{tabular}{lclllllll}\hline\Tstrut\Bstrut
isotope & Abundance (\%) & b(fm)& $g^n_ {pair}$ & $g^p_ {pair}$&  $\Delta_{p}^{th}$ & $\Delta_{p}^{exp}$ & $\Delta_{n}^{th}$& $\Delta_{n}^{exp}$ \\
\hline
$^{156}$Gd  & 20.5  & 2.319&  0.75 & 0.80 &0.960  & 0.961& 1.09 & 1.069\\
$^{158}$Gd  & 24.8  & 2.324&  0.80& 0.77 &  0.881  &0.879 & 1.08& 0.893\\
$^{160}$Gd  & 21.8  & 2.328&  0.81& 0.78 &  0.884  &0.857 & 1.05  &0.831\\
 \hline
\end{tabular}
\end{center}
\label{pairing}
\end{table}
In the pnQRPA calculations the interaction matrix elements were
scaled separately for each multipole  state. In this way the
lowest   excitation energy of each multipole was brought as close
as possible to the experimental energy spectra.
 As an example   in Fig. \ref{fasma} the calculated
 energy spectrum of $^{158}$Tb  together with the experimental one
\cite{tb158} is presented.

\begin{figure}[htb]
\centering
\includegraphics[width=.3\textheight]{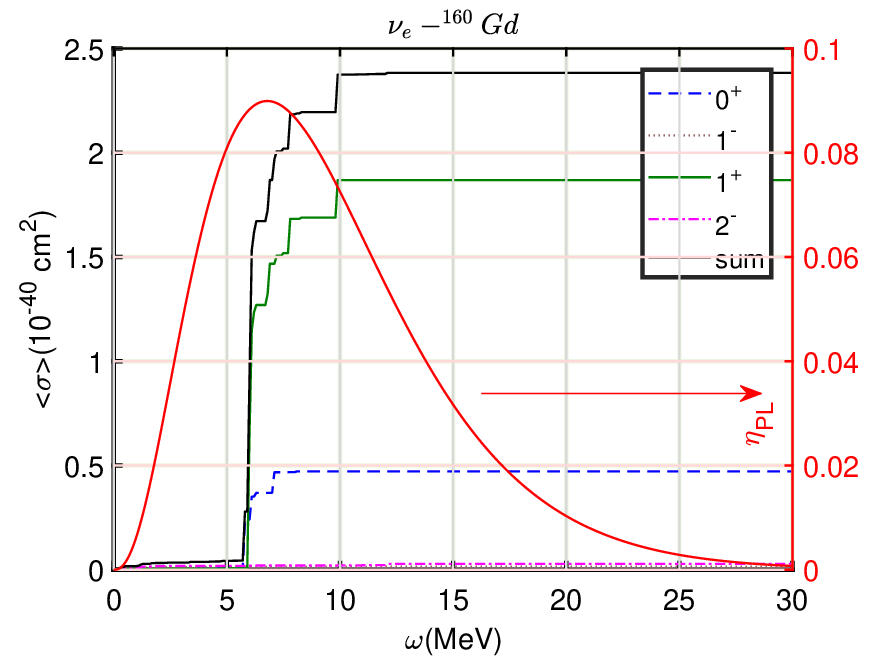}a)
\includegraphics[width=.3\textheight]{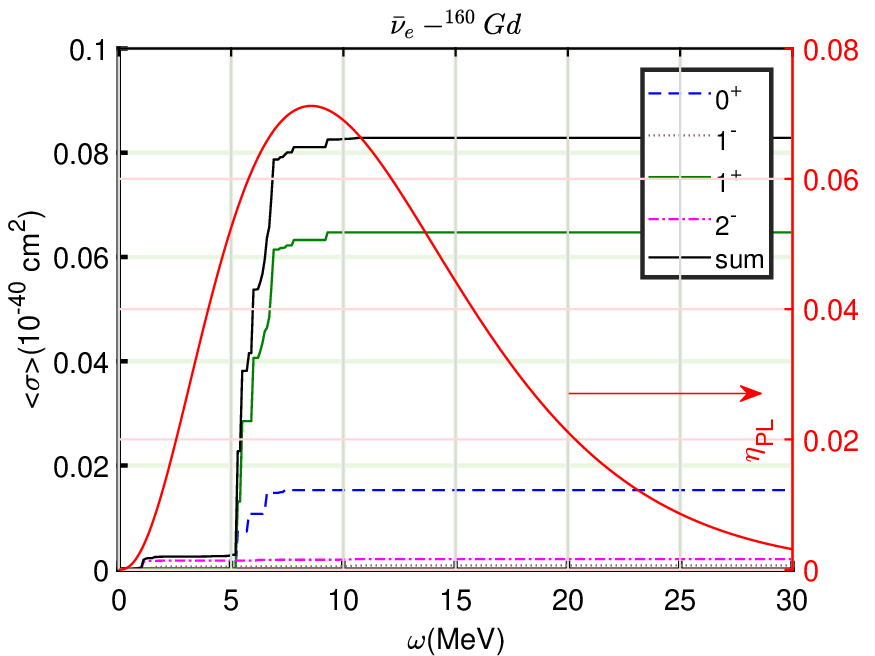}b)
\vspace*{0.3cm} \caption{(Color on line)  Cumulative flux-averaged
cross sections (in units $ 10^{-40}cm^2$) as a function of
excitation energy $\omega$ for the reactions     $^{160}Gd(\nu_e ,
e^-)^{160}Tb$ and $^{160}Gd(\bar\nu_e , e^+)^{160}Eu$.
  Both  multipole state contribution of  $J=0^+,1^+,1^-$
 $2^-$  and  the total sum are presented. The power-law(PL)  distributions $\eta_{PL}$ (red solid line) for $\langle E_{\nu_e}\rangle=9.5$MeV (a)
 and $\langle E_{\bar\nu_e}\rangle=12$MeV (b) with $\alpha=2.5$
 are also displayed.} \label{cumulat}
\end{figure}

In Fig.~\ref{cs1}(a) we present the numerical results of the total
scattering cross section $\sigma(E_{\nu})$ given by
Eq.~(\ref{eq:Sec2_1})  as a function of the incoming neutrino
energy $E_{\nu}$ for the reactions $^{A}Gd(\nu_e , e^-) ^{A}Tb$
and $^{A}Gd(\bar{\nu}_e , e^+) ^{A }Eu$, A=156,158,160
respectively. The Q values ($Q=M(A,Z\pm 1)-M(A,Z))$  of the
reactions are  given in Table \ref{qvalue}.
\begin{table}[htb]
\caption{Q values (in MeV) for the corresponding
neutrino-nucleus interactions.}
\renewcommand{\tabcolsep}{1.0pc} 
\renewcommand{\arraystretch}{2.0} 
\begin{center}
\begin{tabular}{lllllll}\hline
& $\nu_e$-$^{156}$Gd  & $\bar\nu_e$-$^{156}$Gd &
$\nu_e$-$^{158}$Gd  & $\bar\nu_e$-$^{158}$Gd & $\nu_e$-$^{160}$Gd
& $\bar\nu_e$-$^{160}$Gd \\ \Tstrut\Bstrut
 Q (MeV)  & 2.444  & 2.449&  1.219& 3.487 & 0.106& 4.579\\
 \hline
\end{tabular}
\end{center}
\label{qvalue}
\end{table}
The overall cross sections $\sigma(E_{\nu})$ includes a summation
over transitions to all possible final states characterized by
mutipoles up to $J^{\pi}=6^{\pm}$. Here we have considered a
hybrid prescription already used in previous calculations
\cite{Divarijpg2,athar04,Lazauskas}, where Fermi function for
Coulomb correction is used below the energy region  on which both
approaches predict the same values, while EMA is adopted above
this energy region. As it is seen, both the neutrino and
antineutrino cross sections increase strongly with increasing
neutrino energy while the the $\nu_e-$nucleus cross sections are
about an order of magnitude greater than the corresponding
antineutrino ones. For comparison in Fig. \ref{cs1}(a) we also
present the   total cross sections for inverse beta decay, elastic
scattering on electrons and neutral current scattering on oxygen.
For clarity     figure \ref{cs1}(b) plots the energy cross
sections for $\nu_e-Gd$ and  $\bar\nu_e-Gd$ reactions in the
energy region 0 to 15 MeV.

\begin{figure}[htb] \centering
\includegraphics[width=.3\textheight]{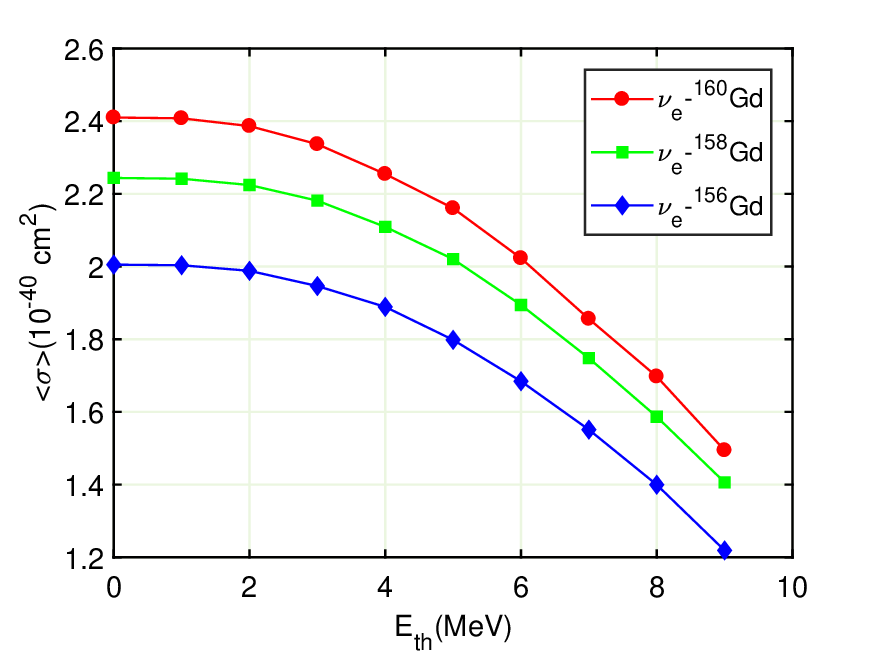}
\caption{(Color on line)  Flux averaged cross sections (in units $
10^{-40}cm^2$) of supernova neutrinos at  $\langle
E_{\nu_e}\rangle=9.5$MeV as a function of energy threshold on the
outgoing electrons.} \label{thres}
\end{figure}

The   flux-averaged supernova-neutrino ($SN-\nu$) cross sections,
broken down by multipoles, appear in  Table \ref{tab:fluxavernu}.
The pinching parameter $\alpha$ has been taken the value $
\alpha=2.5$. As it is seen, at a typical supernova neutrino
(antineutrino) mean energy  $\langle E_{\nu_e}\rangle=9.5$~MeV
($\langle E_{\bar\nu_e}\rangle=12$~MeV) the flux-averaged cross
sections are dominated by the allowed (A) transition moments
$J^{\pi}=0^+,1^+$ contributing about 97\% of the total strength.
The remaining part of the transition strength (3\%) is carried
almost entirely  by the first forbidden (F1) moments
$J^{\pi}=1^-,2^-$ and the second forbidden
(F2):~$J^{\pi}=2^+,3^+$.
 Moreover, in Fig.\ref{cumulat} the
cumulative flux-averaged cross section is illustrated as a
function of the excitation energy $\omega$ for the reactions
$\nu_e-^{160}Gd$ and $\bar\nu_e-^{160}Gd$. As it is seen, the
dominant transitions lie to the energy  region between 5-10 MeV.
The region of maximum discontinuity of the cumulative cross
sections coincides with the maximum multipole contribution of the
$1^+$ states, while, the shape of  neutrino/antineutrino energy
spectrum probes the giant resonance region of the nuclear spectrum
where the cross sections vary quickly.
\begin{table}[tbp]
 \caption{Fraction (in $\%$) of the flux-averaged
cross section associated to states of a given multipolarity with
respect to the total flux-averaged
cross section, i.e. $\langle \protect\sigma \rangle_{J^{\protect\pi%
}}/\langle \protect\sigma \rangle_{tot}$. Results are given for
all positive and negative states having total angular momentum $J$
between 0 and 3. The first column gives the considered neutrino
nucleus reaction and the second one the corresponding mean energy
$\langle E_{\nu}\rangle$. The last column give the total
flux-averaged cross sections in units of $10^{-42}~cm^2$. The
pinching parameter is taken to be $\alpha=2.5$.}
\label{tab:fluxavernu}
\renewcommand{\tabcolsep}{0.3pc} 
\renewcommand{\arraystretch}{1.0} 
\begin{center}
\begin{tabular}{|c|c|c|c|c|c|c|c|c|c|c|}
\hline\Tstrut\Bstrut
 & $\langle E_{\nu}\rangle$(MeV) & $~0^{+}~$& $~1^{+}~$& $~2^{+}~$& $~3^{+}~$&$~0^{-}~$ & $~1^{-}~$& $~2^{-}~$&  $~3^{-}~$ & $\langle \protect\sigma \rangle_{}~($$10^{-42}~cm^2)$ \\
  &  &  &  &  &  &  &  &  &   & $E_{th}=0$ \\
\hline\Tstrut\Bstrut
$\nu_e-^{156}$Gd     &9.5  &15.22   & 81.55   & 0.61  & 0.46  &0.04   &0.37   &1.68  &  0.02   & 200 \\
$\nu_e-^{158}$Gd     & 9.5 &17.46   & 78.86   &0.63   & 0.45  &0.06   &0.54   &1.93  & 0.02    &  224 \\
$\nu_e-^{160}$Gd     & 9.5 &19.58   & 77.50   &0.61   & 0.43  &0.04   &0.60   &1.15  & 0.02    &   241\\
$\bar\nu_e-^{156}$Gd &12   &18.26   &78.44    & 0.50  & 0.36  &0.03   &0.56   &1.79  &0.02    &  13.7  \\
$\bar\nu_e-^{158}$Gd &12   &19.14   &77.46    & 0.58  & 0.38  &0.05   &0.81   &1.51  &0.02   &   9.7 \\
$\bar\nu_e-^{160}$Gd &12   &18.22   &77.26    & 0.59  & 0.38  &0.07   &0.99   &2.45  &0.02   &   8.4 \\
\hline
\end{tabular}%
\end{center}
\end{table}
The above results refer to an ideal detector operating down to
zero threshold $E_{th}=0$.  In the case of non zero threshold the
flux averaged cross sections will be suppressed. This is
demonstrated in Fig. \ref{thres} where   as an example the flux
averaged cross sections for supernova neutrinos at $\langle
E_{\nu_e}\rangle=9.5$  MeV is plotted assuming a threshold
$E_{th}$ on the recoiling electron. As it is seen for an electron
total energy threshold of 5 MeV (energy threshold in SK) the
suppression to the flux flux-averaged cross section $\langle
\sigma\rangle$ is about 9\%.

\begin{figure}[htb]
\begin{center}
\includegraphics[scale=0.4]{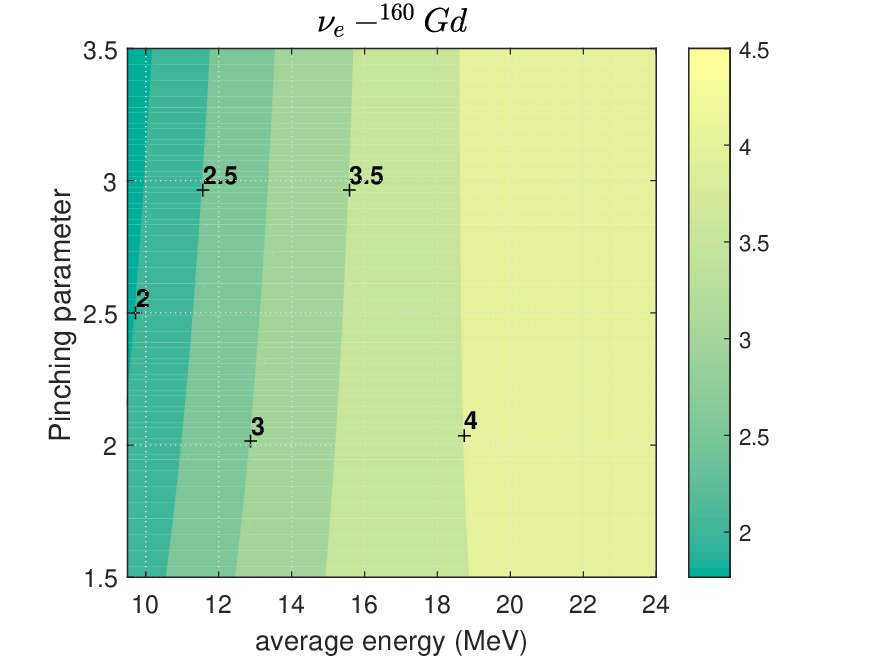}
\includegraphics[scale=0.4]{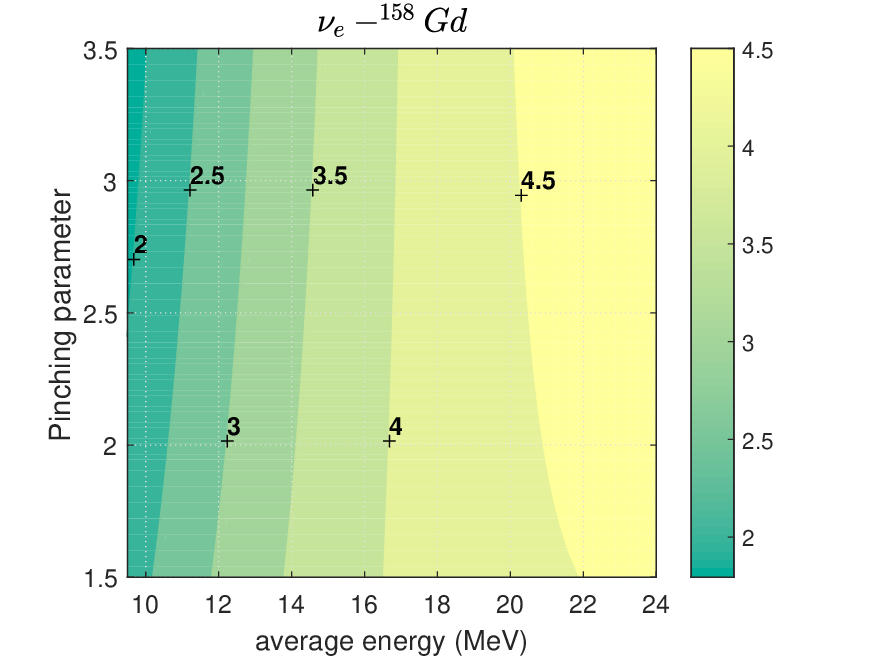}
\includegraphics[scale=0.4]{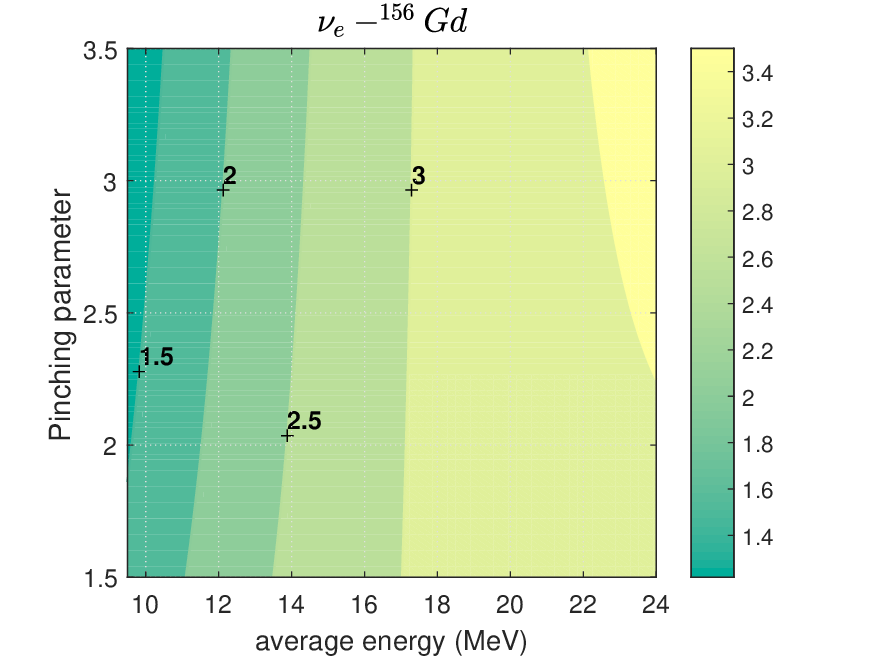}
\\
\includegraphics[scale=0.4]{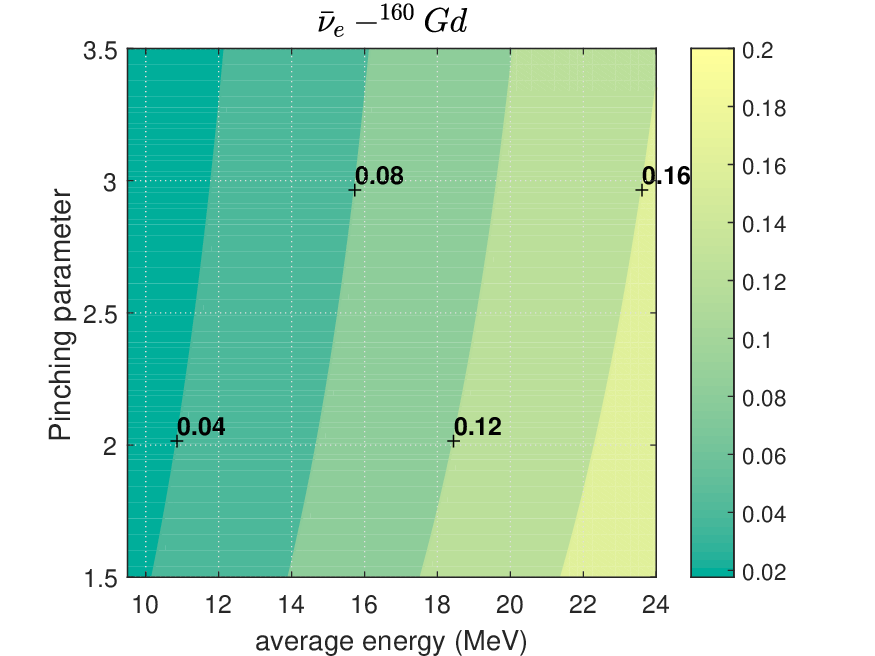}
\includegraphics[scale=0.4]{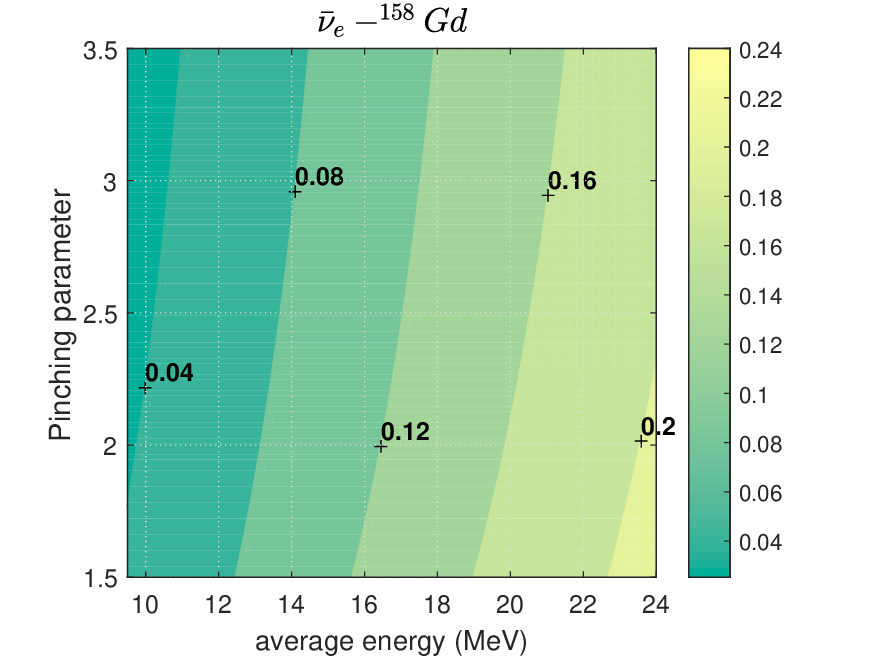}
\includegraphics[scale=0.4]{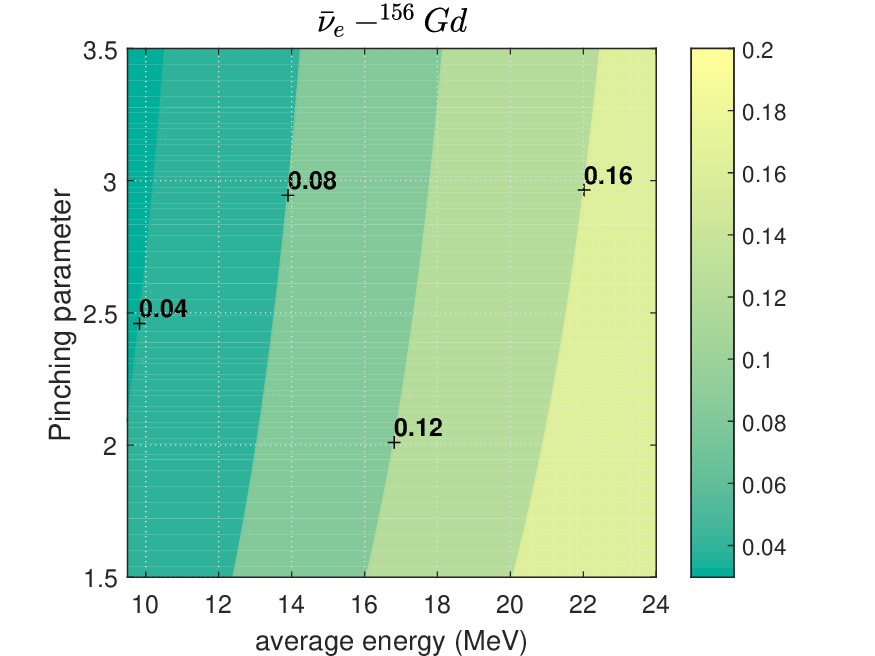}
\caption{(Color on line) These contour plots show the number of
electrons (top panels) and positrons (bottom panels) emitted from
32~tons of Gd consisting of A=156,158 and 160 isotopes. The
contours from left to right in each panel  denotes the increase of
the number of expected events. We have assumed a 5 MeV  detection
threshold.} \label{alphaT}
\end{center}
\end{figure}

Exploiting our predictions for the  total cross sections $\nu-Gd$,
the number of expected    neutrino events are estimated in a WCD
assuming  the addition of  0.1\% (by mass) Gd doping. Thus in the
SK detector where the  fiducial mass of water is 32~ktons the
Gadolinium mass there would  be about $m_t=32$~tons. A supernova
radiates via neutrinos an amount of total energy $3\times10^{53}$
erg in about $10~s$. Assuming an equal partition of energy among
neutrinos, the supernova radiates $N_{\nu_e}=3.0\times 10^{57}$
electron neutrinos and $N_{\bar\nu_e}=2.6\times 10^{57}$ electron
antineutrinos.   The neutrino fluence $\Phi(E_{\nu})$ for
neutrinos integrated over 15 s burst is given by the relation
\begin{equation}
\label{fluxeq} \Phi_i(E_{\nu})=\frac{N_i}{4\pi D^2}
 \eta_{PL}(E_{\nu}), \quad i=\nu_e,\bar\nu_e
\end{equation}
at a distance $D=10$~kpc=$3.1\times 10^{22}$cm. If the mass of the
target material is $m_t$, corresponding  to $N_{t}$ atoms then the
number of expected events   are
\begin{equation}
 N_{event}   =N_{t}\int \Phi_i(E_{\nu})\sigma_i(E_{\nu})dE_{\nu}=
 N_{t}\frac{N_i}{4\pi D^2}\langle \sigma_i\rangle \label{exam1}
\end{equation}
where $\langle \sigma_i\rangle $  the flux-averaged cross
sections.
\begin{table}[htb]
  \caption{Number  of expected events in Super-Kamiokande for a Galactic supernova at a distance of 10~kpc for different values of
    averaged energy. The total energy of the
supernova is assumed to be $3\times 10^{53}$ erg, equally
partitioned  among all flavors (here $\nu_x=\nu_\mu+\nu_\tau$ and
$\bar\nu_x=\bar\nu_\mu+\bar\nu_\tau$). The detection threshold is
taken 0~MeV(events in first parenthesis, 3~MeV(events in second
parenthesis) and 5~MeV(events in third parenthesis). The pinching
parameter $\alpha$ is taken 2.5. The fiducial mass of water that
is being  considered is 32~ktons with  32~tons of Gd.}
\begin{center}
    \begin{tabular}{|l|c|c|c|c|}
        \hline\Tstrut
        Detection channel &    9.5MeV & 12MeV & 15.6MeV  &$N_t(10^{29})$  \\
        \hline\Tstrut
        $ \bar\nu_e+p\rightarrow e^++n $ &(4595.75)(4565.59)(4421.50)&(5914.62)(5900.56)(5823.53)&(7686.76)(7681.19)(7646.82) & 21414\Tstrut \\
        $ \nu_e+e^-\rightarrow \nu_e+e^- $ & (253.49)(169.43)(121.64) &(255.22)(187.84) (147.19) & (256.76)(204.45)(171.64) &85333 \Tstrut\\
         $ \bar\nu_e+e^-\rightarrow \bar\nu_e+e^- $ & (107.51)(47.37)(27.56) & (108.04)(56.64)(36.94) &(108.52)(66.16)(47.66) &85333 \Tstrut  \\
          $ \nu_x+e^-\rightarrow \nu_x+e^- $ & (85.51)(53.10)(36.96) & (85.59)(59.18)(45.07) &(85.64)(64.81)(53.05) & 85333\Bstrut\\
          $ \bar\nu_x+e^-\rightarrow \bar\nu_x+e^- $ &(73.92)(43.41)(29.50) & (73.90)(48.78)(36.33) & (73.87)(53.84)(43.22) &85333 \Bstrut \\
           \hline\Tstrut
       $ \nu_e+^{16}O\rightarrow \nu_e+^{16}O $ & 0.75 & 3.64 & 16.95 & 10667\Bstrut \\
       $ \bar\nu_e+^{16}O\rightarrow \bar\nu_e+^{16}O $ &  0.59 & 2.88 & 13.41& 10667 \Bstrut \\
       \hline
       $  \nu_e+^{160}Gd\rightarrow e^-+^{160}Tb$ &(1.98)(1.97)(1.92)&(2.75)(2.74)(2.68)&(3.58)(3.57)(3.51)&0.26\Tstrut \\
       $  \nu_e+^{158}Gd\rightarrow e^-+^{158}Tb$ &(2.02)(2.01)(1.96)&(2.88)(2.87)(2.82)&(3.81)(3.80)(3.75) & 0.30\Bstrut \\
       $  \nu_e+^{156}Gd\rightarrow e^-+^{156}Tb$ & (1.39)(1.39)(1.37)&(2.06)(2.05)(2.04)&(2.79)(2.78)(2.77) & 0.25 \Bstrut \\
\hline
        $  \nu_e+^{}Gd\rightarrow e^-+^{}Tb$ & (5.39)(5.37)(5.25)&(7.69)(7.65)(7.54)&(10.18)(10.15)(10.03) &0.81\Bstrut \\
 \hline\Tstrut
       $ \bar\nu_e+^{160}Gd\rightarrow  e^++^{160}Eu $ &(0.025)(0.023)(0.023)&(0.047)(0.047)(0.046)&(0.084)(0.085)(0.085) &0.26\Bstrut \\
       $ \bar\nu_e+^{158}Gd\rightarrow  e^++^{158}Eu $ & (0.032)(0.032)(0.032)&(0.061)(0.061)(0.059)&(0.104)(0.104)(0.101)& 0.30\Bstrut \\
       $ \bar\nu_e+^{156}Gd\rightarrow  e^++^{156}Eu $ &(0.037)(0.037)(0.036)&(0.064)(0.064)(0.064)&(0.102)(0.102)(0.102) &0.25\Bstrut \\
       \hline\Tstrut
$ \bar\nu_e+^{}Gd\rightarrow  e^++^{}Eu $ &(0.094)(0.092)(0.090)&(0.171)(0.171)(0.169)&(0.290)(0.291)(0.287)&0.81\Bstrut \\
        \hline
        \end{tabular}\end{center}\label{tab:cost2}
\end{table}
In Fig.~\ref{alphaT} a contour plot is used to display the number
of expected events   for the reactions $Gd(\nu_e,e^-)Tb$ and
$Gd(\bar\nu_e,e^+)Eu$ respectively, with various parameterizations
of power-law spectra. As it is seen, within a window of 10-18 MeV
the number of events depends  weakly on the pinching parameter
$\alpha$. However, at energies out of this region, the number of
events increases   faster.

\begin{figure}[htb] \begin{center}
\includegraphics[width=.3\textheight]{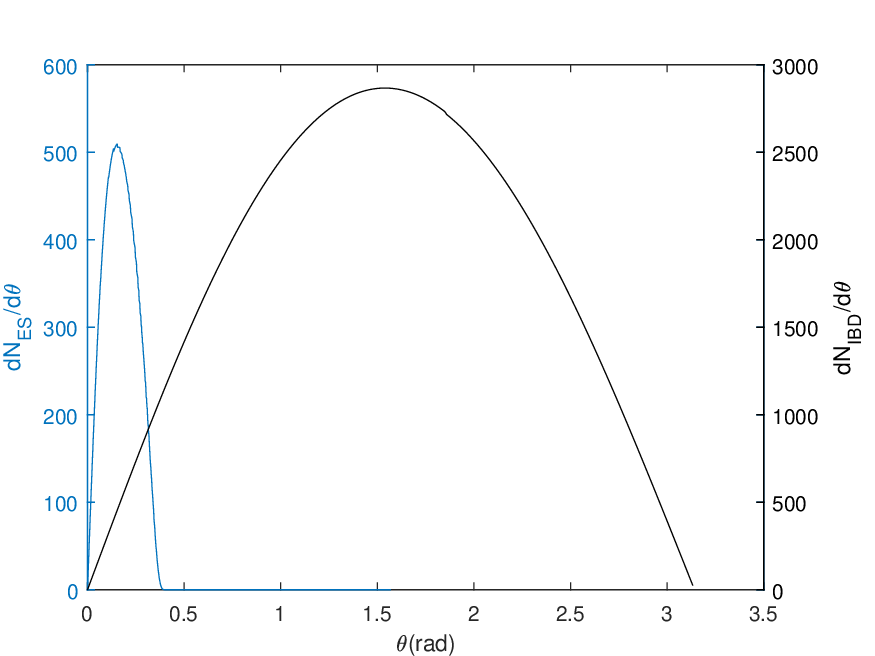}a)
\includegraphics[width=.3\textheight]{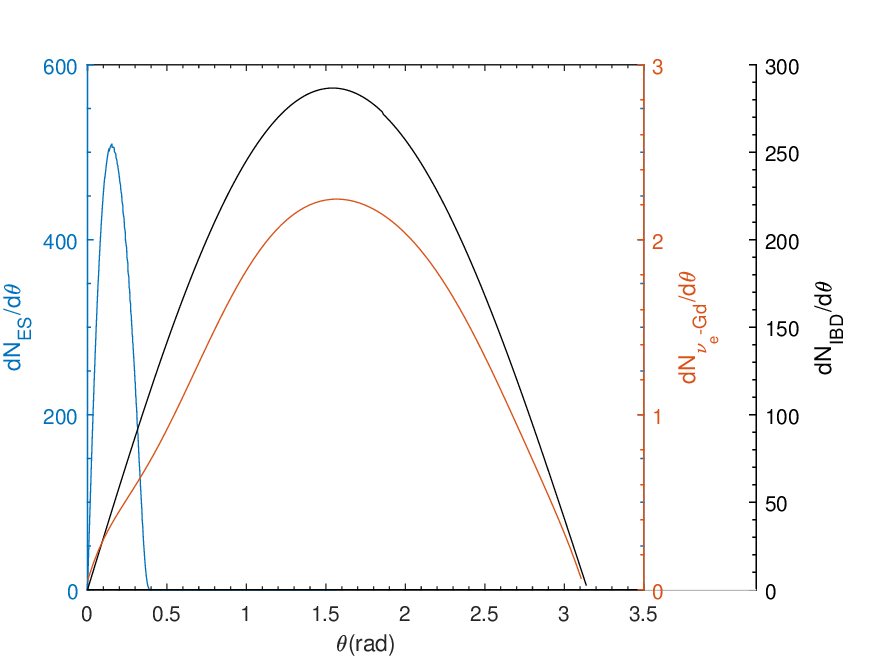}b)
\caption{(Color on line) Angular distributions   of events for ES,
IBD and $\nu_e-$Gd   as a function of the scattering angle
$\theta$ without(left) or with Gd(right). We take $\langle
E_{\nu_e}\rangle=9.5$MeV and $\langle E_{\bar\nu_e}\rangle=12$MeV.
We assume 32~tons of Gd. The energy threshold is taken 5~MeV.}
\label{angular}
\end{center}
\end{figure}

Next  the number of expected events in SK detector for a Galactic
Supernova at 10kpc and for different values of the neutrino
average energy are estimated in Table \ref{tab:cost2}. We consider
32~ktons fiducial mass assuming a 100\% tagging efficiency on
expected events above the detection  threshold. For comparison the
detectable channels $\bar\nu_e+p$ (IBD), the elastic scattering
$\nu_e+e^-$ (ES) as well as the neutral current scattering
$\nu_e+^{16}O$ (OS) on oxygen are also calculated
\cite{Tomas1,Strumia2003}. Furthermore, in Fig. \ref{angular} it
is shown the angular distributions of events as a function of
scattering angle for the various detection channels.
 From the table, it is clear that the largest number of events will
be due to the IBD which is almost isotropic  \cite{Vogel:1999zy},
while the ES events spread out in a cone of about $20^\circ$
\cite{Nakahata:1998pz} that points towards the neutrino direction
(see Fig. \ref{angular}a). Thus positrons from IBD and electrons
from ES  can be statistically distinguished by reducing the IBD
background to the portion of the solid angle in which it overlaps
to the ES signal. Beacom and Vagins \cite{Beacom:2003nk} suggest
that  with    0.1\%  Gd  added to  SK, $\sim 90\%$ of the IBD
events could be tagged. The remaining IBD events as well as the
$\bar\nu_e$ absorption events on $^{16}$O can be statistically
subtracted from the remaining signal. As it is clear from Table
\ref{tab:cost2} the $\nu_e$ interactions on electrons are the
largest in number among electron scattering interactions.
Moreover, the $\nu_e-$Gd charged current interactions as well as
   ES  depend weakly on the average energy of the incoming
neutrino. As it seen is from figure \ref{cumulat}, $\nu_e-$Gd
events could be identified by the expected gamma lines in the
energy window 5-10\hspace{2pt}MeV. The $\bar\nu_e-$Gd interactions
are quite small and are hidden by the large IBD interactions on
free protons. As regards the OS signal \cite{Rosso2018}, it is
expected to be within  $4 \div 9$~MeV, gamma lines cover the
energy window $\approx 5.3\div 7.3$~MeV . In this region it can
not be disentangled from the many more IBD and ES background
events.
The main background for ES  and $\nu_e-$Gd interactions are the
IBD events. Some of these numerous events can be removed using an
angular cut but they still pose a formidable background (see Fig.~
\ref{angular}a). Adding Gd to SK the inverse beta background will
decrease about 90\% (see  Fig.~\ref{angular}b). This could improve
the detection prospects of  ES channel which is strongly forward
peaked. The ability to cleanly isolate  the dominant IBD events
would be extremely important for studying the remaining reactions
$\nu_e-$Gd that lead to gamma emission. If $\nu_e-$Gd events could
be isolated either by gamma rays identification
  or by the determination of probable delayed beta
decays, they might have some advantages due  to the low thresholds
(though low yields). Recently
 a new method was proposed \cite{Dongol} to introduce  Gd-ions
 in WCDs, based to release of Gd-ions from custom designed glasses
 like those used for photomultiplier tube  glass systems.
This controlled Gd-ion release from a custom glass in the form of
beads or powders may help in future WCDs  to enhance neutrino
detection.

\section{Conclusions}

The addition of   Gadolinium (Gd) salt in the Water Cherenkov
Detectors   enhances the sensitivity to neutrino detection. In
this work we have computed the cross sections for charged current
neutrino and antineutrino scattering off the even A=156-160 (most
abundant) Gd isotopes for   energies relevant to supernova
neutrinos.   The neutrino induced transitions to excited nuclear
states are computed in the framework of  pnQRPA.  The nuclear
responses of the Gd  isotopes for SN detection have   been studied
assuming  a two-parameter quasi-thermal power law distribution.
Our results    show that the greatest part of responses comes from
the excitation energy region $\omega < 20~MeV$. The neutrino-Gd
channel have also been compared with three other channels, namely
inverse beta decay, elastic scattering on electrons and neutral
current scattering on oxygen. We tried to look at the angular
dependence of the $\nu_e-$Gd interaction signal in the SK detector
with fiducial mass 32~ktons of water and 32~tons Gd doping. The
problem is the background of events from inverse beta channel.
This background can be reduced for elastic scattering on electrons
using an angular cut. The number of $\nu_e-$Gd events are
increasingly backward peaked and are about 80 times smaller than
those of inverse beta events. It would be also interesting to
investigate cross sections for charged current neutrino scattering
off the odd $^{155,157}$Gd isotopes. Detailed numerical  results
will be presented  in a forthcoming paper. The sensitivity of the
detectors, which is pivotal to the success of Water Cherenkov
Detectors can be achieved including additives, such as Gd in
water, something that is more financially sound and a less risky
option than, either building a larger water tank or varying the
size of photomultiplier tubes. The ability to a well understood
reducible backgrounds above detector threshold is extremely
important for studying charged current signals from supernova.

\bibliography{TEX}
\end{document}